**HRPv2: an automated and enhanced method for full-length homology-based R-gene prediction**

Giuseppe Andolfo[1,*], Thomas Holzweber[2], Juliane C. Dohm[2] and Heinz Himmelbauer[2,*]

[1]Department of Agricultural Sciences, University of Naples ‘Federico II’, Via Università 100, 80055 Portici (Naples), Italy

[2]Institute of Computational Biology, Department of Biotechnology, BOKU University, Muthgasse 18, 1190 Vienna, Austria

*To whom correspondence should be addressed.

giuseppe.andolfo@unina.it

heinz.himmelbauer@boku.ac.at

**Abstract**

**Motivation**: Plant disease resistance genes, particularly those encoding NB-LRR proteins, are important targets for crop improvement. Proteome-based domain or motif searches can only identify NB-LRRs among existing gene models, meaning they cannot recover loci that have been missed or incorrectly predicted by the reference annotation. The full-length, homology-based R-gene prediction (HRP) method circumvents this issue by reconstructing gene models directly on the genome. However, the original implementation of this method requires several separate phases of classification, comparison and filtering operations.

**Results**: HRPv2 is an automated, enhanced version of the original strategy. The labour-intensive, step-by-step curation process used in HRP has been replaced by a reproducible filtering framework. Enhancing two steps of the homology search process has enabled HRPv2 to better account for the specific NB-LRR variability of the genome. Performance validation confirmed that the number of full-length NB-LRRs annotated in both the automatically predicted gene set of the respective genome assembly and the final NB-LRR repertoire has increased in HRPv2 compared to HRP.

**Availability and implementation**: HRPv2 and its associated documentation and reproducible test data are available at https://github.com/AndolfoG/HRPv2. Detailed installation and dependency information is provided in the repository README. A version-pinned Conda package has also been developed and locally validated to provide a reproducible execution environment.

**Supplementary data** are available at https://doi.org/10.6084/m9.figshare.34085769.

## 1 Introduction

Plant NB-LRR genes tend to occur in rapidly evolving clusters and are frequently fragmented, fused or missing in gene sets after general-purpose gene prediction. Searches performed against an annotated proteome alone inherit these shortcomings. Therefore, motif- and domain-based proteome screens are constrained by the completeness of the initial annotation (Andolfo et al., 2022).

HRP overcame this limitation by using full-length NB-LRR homology to search the reference genome sequence and reconstruct a more comprehensive set of candidate gene models supported by existing annotations. The original study demonstrated that this genome-based strategy recovered substantially more NB-LRR genes than searches restricted to predicted proteins and their domains or motifs (Andolfo et al., 2022).

The main practical limitation of HRP was not the prediction strategy itself, but the extensive curation required after the prediction had been made. Candidate models had to be examined in successive manual steps. Short and excessively long models were removed; protein-domain architectures were checked; non-NB-LRR predictions were excluded; overlapping alternatives were compared on the same strand; and atypical fusion-like predictions were inspected. These operations ensured high-quality output but were slow and tedious, and were subject to inconsistent decision-making. Furthermore, the first step of the homology search only consulted two InterProScan databases (Pfam and SUPERFAMILY) for screening the typical NB-LRR domain (Jones et al., 2014). Additionally, retrieving additional R-genes by MEME/MAST was restricted to a non-directed motif search, disregarding the structural features of the NB protein domain (Bailey, et al., 2006).

Here, we present an automated, enhanced version of the full-length, homology-based R-gene prediction pipeline (HRPv2), which uses a reproducible filtering framework to comprehensively identify the full NB-LRR gene repertoire within genome assemblies. We compared the performance of HRPv2 with the manually curated HRP annotation for the tomato (*Solanum lycopersicum*) genome (Andolfo et al., 2022) and show that HRPv2 outperforms HRP in terms of number and completeness of resulting NB-LRR gene models.

## 2 Methods

### 2.1 NB-LRR prediction framework

The Homology-based R-gene Prediction pipeline (HRP) employs a two-level homology-search strategy (Andolfo et al., 2022). First, a protein motif/domain-based search identifies full-length NB-LRR proteins within an automatically predicted gene set. These genome-specific representatives are then used as queries against the corresponding genome assembly to recover paralogous NB-LRR loci that were missing, fragmented, or incorrectly modelled during automated annotation.

HRPv2 retains this architecture while introducing three major methodological advances. First, it implements the entire procedure as a unified command-line workflow, automating input validation, protein annotation, query selection, gene prediction, architecture classification, redundancy filtering, and construction of the final NB-LRR repertoire. This integration minimizes manual intervention and provides traceable intermediate outputs, logs, metadata, and diagnostic reports. Second, HRPv2 substantially expands the protein domain-based search underlying the first-level of homology search. Rather than relying exclusively on Pfam and SUPERFAMILY, it integrates complementary evidence from all InterProScan member databases and prediction methods (Jones et al., 2014).

Third, HRPv2 introduces a genome-specific NB domain rescue module that refines query selection before the second-step of homology search. High-confidence PF00931 regions detected by InterProScan are aligned and used to construct a profile hidden Markov model with HMMER and conserved motif profiles by MEME with default settings except for the following parameters -nmotifs 8 -minw 6 -maxw 10 (Bailey, et al., 2006). MEME learns eight profiles from four conserved NB-ARC anchors — P-loop, Kinase-2, GLPL, and MHD — and the intervening RNBS-A, RNBS-B, RNBS-C, and RNBS-D regions (Meyers et al., 1999). Additional NB candidates are retained only when statistically supported HMMER matches are corroborated by at least three motifs detected by MAST with default settings, in a biologically consistent order. Rescued NB regions enter the full-length query set only in the presence of independent downstream LRR evidence. Consequently, HRPv2 provides GenBlastG with a query set that more accurately represents the species-specific diversity of the NB-LRR repertoire (She, et al., 2011).

### 2.2 HRPv2 validation

To assess the performance of the automated workflow, HRPv2 was benchmarked against the tomato NB-LRR repertoire previously generated with HRP (Andolfo et al., 2022). HRPv2 is being run on the *S. lycopersicum* Heinz SL2.4 genome with the matching proteome and structural annotation used for the previous study. Using the same reference versions minimizes assembly-driven differences and permits a direct comparison with the published NB-LRR gene catalogue.

The first-level homology search identified 173 and 194 full-length NB-LRR queries with HRP and HRPv2, respectively, while the second-level genomic search expanded the corresponding full-length repertoires to 231 and 260 loci (Table 1; Supplementary Table 1). Evaluation of accuracy in the prediction of gene models showed that HRPv2 recovered all 231 full-length loci identified by HRP, with a mean locus coverage of approximately 90%, and intercepted 102 of the 132 previously reported partial NB-LRR loci (Supplementary Table 1). Of the remaining 30 partial NB-LRR genes identified by HRP, 27 were not confirmed by HRPv2, despite being predicted by the second-level homology search, because they did not meet the improved annotation criteria implemented in HRPv2. The annotated leucine-rich regions cannot be unequivocally attributed to the NB-LRR gene family. Similarly, conserved NB motifs do not provide sufficient evidence to identify new partial genes (Supplementary Table 2). Finally, only three loci identified by HRP were not predicted by HRPv2 (Supplementary Table 3).

**Table 1**. Number of tomato NB-LRR genes as identified by HRP and HRPv2 in *Solanum lycopersicum* Heinz-1706 v2.4.

| | **first-level of homolgy search** | | **second-level of homology search** | |
|---|---|---|---|---|
| **Protein domain[A]** | **HRP[B]** | **HRPv2** | **HRP[B]** | **HRPv2** |
| CC-NB-LRR | 152 | 171 | 198 | 225 |
| TIR-NB-LRR | 19 | 21 | 31 | 33 |
| RPW8-NB-LRR | 2 | 2 | 2 | 2 |
| **Total full-length** | 173 | 194 | 231 (231) | 260 |
| **Total partial** | 87 | 171 | 132 (102) | 127 |
| **Total** | 260 | 365 | 363 (333) | 387 |

[A]CC: coiled coil; LRR: leucine-rich repeat; NB: nucleotide binding; RPW8: resistance to powdery mildew 8; TIR: Toll/interleukin receptor. [B]Number of confirmed NB-LRR genes previously identified by HRP approach in brackets.

## 3 Availability, reproducibility and use

HRPv2 is implemented in Python and is publicly available through the GitHub repository (https://github.com/AndolfoG/HRPv2), together with documentation, installation guidelines, and example datasets. Two installation routes are available on GitHub: manual installation using the repository version of HRPv2.py, and Conda installation using the versioned release and the recipe in the conda/ directory. The README describes the workflow architecture, command-line parameters, input requirements, and intermediate and final outputs, whereas the accompanying Installation guidelines provide step-by-step instructions for configuring the required software, dependencies, and execution environments. To ensure reproducibility and traceability, HRPv2 retains intermediate results, execution logs, run manifests, parameter settings, software-version information, and provenance metadata. The validated package is intended for distribution through Bioconda, with package metadata and installation commands exposed through Anaconda.org, thereby enabling reproducible installation with Conda-compatible package managers.

## 4 Discussion

HRPv2 generates a comprehensive, reproducible, and traceable catalogue of full-length and partial NB-LRR loci. This workflow recovers loci that are missing, incomplete, fused or otherwise incorrectly represented in reference gene annotations. Automated comparison of overlapping gene models reduces the inconsistencies that may arise from manual curation and facilitates the reproducible resolution of fragmented, redundant and fusion-like predictions. Preserving intermediate outputs, parameters, software versions, execution logs and provenance metadata enables individual predictions to be inspected and reproduced independently.

Benchmarking against the original, manually curated strategy (Andolfo et al., 2022) showed that the original approach was not compromised by automation. HRPv2 recovered all 231 full-length loci reported by HRP, while expanding the final full-length repertoire to 260 loci. The expanded NB-LRR domain search, based on the genome-specific NB domain rescue module, improve the representation of the analysed genome's intrinsic diversity before genomic homology searches are performed (Andolfo et al., 2022). HRPv2 retains substantial sensitivity towards structurally incomplete candidates, as demonstrated by the identification of 127 partial NB-LRRs. Identifying fragmented genes is particularly important because structurally incomplete loci may be overlooked sources of functional disease resistance, as Tu et al. (2026) recently demonstrated. Furthermore, HRPv2 employs more selective classification and filtering criteria, excluding weakly supported partial predictions and avoiding false positives.

These advances make curated NB-LRR repertoires accessible more quickly for disease-resistance research and crop improvement. The resulting gene models support the identification of candidate loci associated with resistant phenotypes, comparative analyses across cultivars and wild relatives, marker development for introgression and selection, and allele mining studies (Andolfo et al., 2022). Accurate discrimination between intact receptors and partial loci is also important for prioritising candidates for functional validation, resistance-gene stacking, and genome-editing strategies.

## 5 Conclusion

In conclusion, HRPv2 is an automated and reproducible tool for the comprehensive discovery and structural classification of full-length and partial NB-LRR loci. By improving the recovery of resistance-gene candidates, HRPv2 facilitates the exploration of natural NB-LRR diversity and provides a robust foundation for comparative genomics, functional validation and the development of crop cultivars with enhanced disease resistance.

## Funding

This work was supported by the RESISTOM projects (Grant numbers E53C25003250001), which are part of the Italian Science Fund (FIS 3 2024-2025) program supported by the Ministry of Universities and Research (MUR). Support to T.H. was provided by the Doctoral School AgriGenomics of BOKU University.

**Conflict of interest**

The authors declare no conflict of interest.

**Data availability**

Information on how to run the HRPv2 pipeline is available on GitHub (https://github.com/AndolfoG/HRPv2.git). Information on input genes and gene predictions can be found at the Solanaceae Genomics Network web page (https://solgenomics.net/) and in the online supplementary material which is part of this article.

**Author contributions**

GA, HH and JCD conceived the study, GA developed the software and analyzed the data. TH tested and evaluated the software implementation. GA wrote the first manuscript draft. All authors finalized the manuscript draft. All authors approved the final manuscript.